\documentclass[floats,aps,showpacs,twocolumn,superscriptaddress]{revtex4-1}
\usepackage[dvips]{graphicx}
\usepackage{amsfonts}
\usepackage{amsmath,amssymb}
\usepackage{dsfont}
\usepackage{placeins}
\usepackage{afterpage}
\usepackage{flafter}
\usepackage[usenames,dvipsnames]{color}
\usepackage{cancel}

\newcommand{\act}{J}
\newcommand{\ang}{\vartheta}

\newcommand{\unpact}{I}
\newcommand{\unpang}{\theta}

\newcommand{\qU}{\bar{q}}
\newcommand{\pU}{\bar{p}}

\newcommand{\angU}{\bar{\ang}}
\newcommand{\actU}{\bar{\act}}

\newcommand{\omegaU}{\bar{\omega}}

\newcommand{\ttilde}{\tilde{t}}

\newcommand{\K}{\kappa}
\newcommand{\qfix}{q^{*}}
\newcommand{\pfix}{p^{*}}

\newcommand{\crvparam}{\lambda}

\newcommand{\diff}{\text{d}}

\newcommand{\HrsActAng}{\mathcal{H}_{r:s}}
\newcommand{\Hrs}{H_{r:s}}

\newcommand{\Hnod}{\mathcal{H}_{0}}
\newcommand{\V}{\mathcal{V}}

\newcommand{\Hpend}{\mathcal{H}_{r:s}^\text{pend}}

\newcommand{\Vrs}{V_{r:s}}
\newcommand{\Irs}{I_{r:s}}
\newcommand{\Mrs}{M_{r:s}}
\newcommand{\monodromy}{\bar{\mathcal{M}}_{r:s}}
\newcommand{\signdisp}{\mu}

\newcommand{\Ors}{\Omega_{r:s}}

\newcommand{\Ntori}{\tau_{\text{max}}}
\newcommand{\Nitersteps}{\ell_{\text{max}}}
\newcommand{\Nmaxorder}{\mathcal{K}}
\newcommand{\Nparams}{\alpha}
\newcommand{\Nsamplepoints}{\mathcal{N}}

\newcommand{\afam}{\mathbf{a}}
\newcommand{\bfzero}{\boldsymbol{0}}

\newcommand{\Lq}{\mathcal{L}_{q}}
\newcommand{\Lp}{\mathcal{L}_{p}}
\newcommand{\Nq}{\mathcal{N}_{q}}
\newcommand{\Np}{\mathcal{N}_{p}}

\newcommand{\x}{\mathbf{x}}

\newcommand{\xU}{\bar{\x}}

\newcommand{\Ars}{A_{r:s}}
\newcommand{\ArsInner}{A_1}

\newcommand{\ArsU}{\bar{A}_{r:s}}
\newcommand{\ArsInnerU}{\bar{A}_1}

\newcommand{\NormalFormCitations}{\cite{Chi1979, Alm88, LebMou1999, LoeBaeKetSch2010, BroSchUll2001, BroSchUll2002, DeuMouSch2013}}
\newcommand{\KAMCitations}{\cite{Kol1954InsertedCitation, Arn1963b, Arn1963, Mos1962} }

\newcommand{\insertfigure}[4]{
\begin{figure}[#1]
  \begin{center}
    \includegraphics{#2}%
    \caption{#3}
    \label{#4}
  \end{center}
\end{figure}
}

\newcommand{\insertbigfigure}[4]{
\begin{figure*}[#1]
  \begin{center}
     \includegraphics{#2}
     \caption{#3}
     \label{#4}
  \end{center}
\end{figure*}
}

\begin{document}

\title{Integrable approximation of regular regions with a nonlinear resonance chain}

\author{Julius Kullig}
\affiliation{Technische Universit\"at Dresden, Institut f\"ur Theoretische
             Physik and Center for Dynamics, 01062 Dresden, Germany}
\affiliation{Max-Planck-Institut f\"ur Physik komplexer Systeme, N\"othnitzer
Stra\ss{}e 38, 01187 Dresden, Germany}
\affiliation{Institut f\"ur Theoretische Physik, Universit\"at Magdeburg, Postfach 4120,
39016 Magdeburg, Germany}

\author{Clemens L\"obner}
\affiliation{Technische Universit\"at Dresden, Institut f\"ur Theoretische
             Physik and Center for Dynamics, 01062 Dresden, Germany}
\affiliation{Max-Planck-Institut f\"ur Physik komplexer Systeme,
N\"othnitzer Stra\ss{}e 38, 01187 Dresden, Germany}

\author{Normann Mertig}
\affiliation{Technische Universit\"at Dresden, Institut f\"ur Theoretische
             Physik and Center for Dynamics, 01062 Dresden, Germany}
\affiliation{Max-Planck-Institut f\"ur Physik komplexer Systeme, N\"othnitzer
Stra\ss{}e 38, 01187 Dresden, Germany}
\affiliation{Department of Physics, Tokyo Metropolitan University, Minami-Osawa, Hachioji 192-0397, Japan}

\author{Arnd B\"acker}\
\affiliation{Technische Universit\"at Dresden, Institut f\"ur Theoretische
             Physik and Center for Dynamics, 01062 Dresden, Germany}
\affiliation{Max-Planck-Institut f\"ur Physik komplexer Systeme, N\"othnitzer
Stra\ss{}e 38, 01187 Dresden, Germany}

\author{Roland Ketzmerick}
\affiliation{Technische Universit\"at Dresden, Institut f\"ur Theoretische
             Physik and Center for Dynamics, 01062 Dresden, Germany}
\affiliation{Max-Planck-Institut f\"ur Physik komplexer Systeme, N\"othnitzer
Stra\ss{}e 38, 01187 Dresden, Germany}

\date{\today}

\begin{abstract}

Generic Hamiltonian systems have a mixed phase space where regions of regular
and chaotic motion coexist. We present a method for constructing an
integrable approximation to such regular phase-space regions
including a nonlinear resonance chain.
This approach generalizes the recently introduced iterative canonical transformation method. 
In the first step of the method a normal-form Hamiltonian with a resonance chain
is adapted such that actions and frequencies match with those of the non-integrable system.
In the second step a sequence of canonical transformations
is applied to the integrable approximation to match the shape of regular tori.
We demonstrate the method for the generic standard map at various parameters.

\end{abstract}

\maketitle
\noindent

\section{Introduction}

Hamiltonian systems are an important class of dynamical systems having
particular relevance for physical applications,
e.\,g., in celestial mechanics, accelerator dynamics,
or mesoscopic and molecular physics.
A special case of Hamiltonian systems are integrable systems, where
the dynamics is restricted to invariant tori in phase space.
The other extreme is given by fully chaotic systems, where the dynamics shows
sensitive dependence on the initial conditions and explores the whole phase
space.

Generic Hamiltonian systems, however, have a mixed phase space where regions of
regular and chaotic motion coexist \cite{MarMey74,LicLie1992,Chi1979}.
This is illustrated using the example of the
standard map in
Fig.~\ref{fig:Mixed_Phase_Space_Intro}(a):
Here, according to the Kolmogorov--Arnold--Moser (KAM) theorem \KAMCitations, a set of regular
tori (lines) forms a regular phase-space region.
As predicted by the Poincar{\'e}--Birkhoff theorem \cite{Poi1912,Bir1913},
these tori are interspersed with nonlinear resonance chains leading to
a rich self-similar structure.
The regular region is embedded in a
phase-space region of chaotic motion (dots).

Constructing integrable approximations to regular phase-space regions is helpful or even necessary for many problems,
e.\,g. for toroidal magnetic devices \cite{HudDew1998},
diffusion in random maps \cite{BazSibTurVai1992,BazSibTur1997,KruKetKan2012},
Arnold diffusion \cite{Arn1964,Cin2002}, or regular-to-chaotic quantum tunneling
\cite{BaeKetLoeSch2008,LoeBaeKetSch2010,BaeKetLoe2010,MerLoeBaeKetShu2013}.
Such an integrable Hamiltonian system
should mimic the dynamics inside the regular phase-space region
as closely as possible.
There are situations where it is essential to include a nonlinear resonance chain into the integrable approximation. 
Here our main motivation is the description of resonance assisted tunneling \cite{BroSchUll2002} 
using complex paths \cite{MerLoeBaeKetShu2013}
and the fictitious integrable system approach \cite{LoeBaeKetSch2010} without perturbation theory.
%% This is the case, e.\,g., for predicting resonance assisted tunneling rates using complex paths.

Up to now, integrable approximations can be provided for near-integrable
systems, e.\,g., by using classical perturbation theory based on
Lie-transforms \cite{LicLie1992,Dep1969,Car1981,BroSchUll2002},
normal-form techniques
\cite{Bir1927,Gus1966,Mey1974,SchWaaWig2006,LebMou1999}, or the
Campbell--Baker--Hausdorff formula \cite{Sch1988,Sok1986,Yos1993}. Also for the more
challenging case of generic non-integrable systems with a mixed phase space,
there are methods available to provide integrable approximations to the regular
phase-space region \cite{BaeKetLoe2010,LoeLoeBaeKet2013}.
Particularly flexible is the recently introduced iterative canonical transformation method \cite{LoeLoeBaeKet2013}
as it independently accounts for the frequencies and the shape of regular tori
and is applicable to higher dimensions.
However, in the generic case, none of these methods is so far capable of producing an integrable
approximation which includes a nonlinear
resonance chain.
\insertfigure{tb}{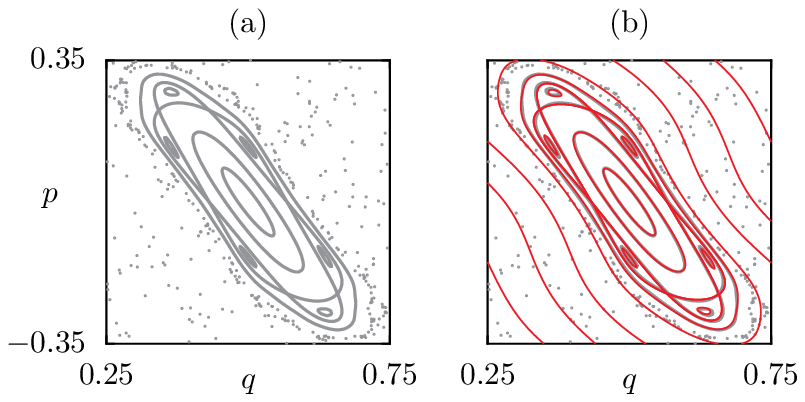}{(color online) (a) Phase space of the standard map, Eq.~\eqref{eq:SMap},
at $\K=3.4$, with regular orbits (gray lines) and chaotic orbits (gray dots)
and (b) its integrable approximation (thin red lines).}{fig:Mixed_Phase_Space_Intro}

In this paper we present a method for constructing an integrable approximation
to a regular phase-space region and one nonlinear resonance chain.
This is achieved by choosing a normal-form Hamiltonian with a resonance chain
% somehow we need a backslash space here, otherwise it wont work
\NormalFormCitations\ as the starting point of the iterative canonical
transformation method of Ref.~\cite{LoeLoeBaeKet2013}.
To illustrate the method, we apply it to the generic standard map,
giving, e.\,g., the integrable approximation of Fig.~\ref{fig:Mixed_Phase_Space_Intro}(b).

The paper is organized as follows:\
In Sec.~\ref{sec:ExampleSystem} we discuss the phase-space structure of a resonance chain
using the example of the standard map.
In Sec.~\ref{sec:ICTM} we present the method for constructing an integrable
approximation to a regular phase-space region and one nonlinear
resonance chain.
In Sec.~\ref{sec:ICTMStd} we apply the method to the standard map.
In Sec.~\ref{sec:Summary} we give a summary and outlook.

\section{Example system with a resonance}
  \label{sec:ExampleSystem}
The construction of integrable approximations described in this paper
applies to time-periodically driven Hamiltonian systems with one degree of freedom.
These systems obey Hamilton's equations of motion,
\begin{subequations}
  \label{eq:HamiltonsEquations}
  \begin{align}
	\dot{q} =&\, \frac{\partial H(q,p,\ttilde)}{\partial p}, \\
	\dot{p} =&\, -\frac{\partial H(q,p,\ttilde)}{\partial q},
  \end{align}
\end{subequations} for position $q$ and momentum $p$.
Considering the corresponding trajectories stroboscopically at times $\ttilde=tT$ with $t\in\mathbb{Z}$,
that are multiples of the external driving period $T$,
gives a symplectic map $U$,
\begin{align}
  \label{eq:Map}
  (q_{t+1}, p_{t+1}) = U(q_t, p_t),
\end{align}
for the evolution of the point $(q_t, p_t)$ to
$(q_{t+1}, p_{t+1})$ in phase space.

The paradigmatic example of such a system is the standard map \cite{Chi1979}
\begin{subequations}
  \label{eq:SMap}
  \begin{align}
	q_{t+1} =&\, q_t + p_t, \\
	p_{t+1} =&\, p_t + \frac{\K}{2\pi} \sin[2\pi (q_t + p_t)],
  \end{align}
\end{subequations}
which we consider for $(q, p) \in [0,1[\times[-0.5, 0.5[$ with periodic boundary conditions.
In this paper we focus on $\K=3.4$. Here the standard map has an elliptic
fixed point at $(\qfix, \pfix) = (0.5, 0)$ which is surrounded by a large
regular phase-space region embedded in a chaotic phase-space region, see
Fig.~\ref{fig:SMap}.

\insertfigure{tb}{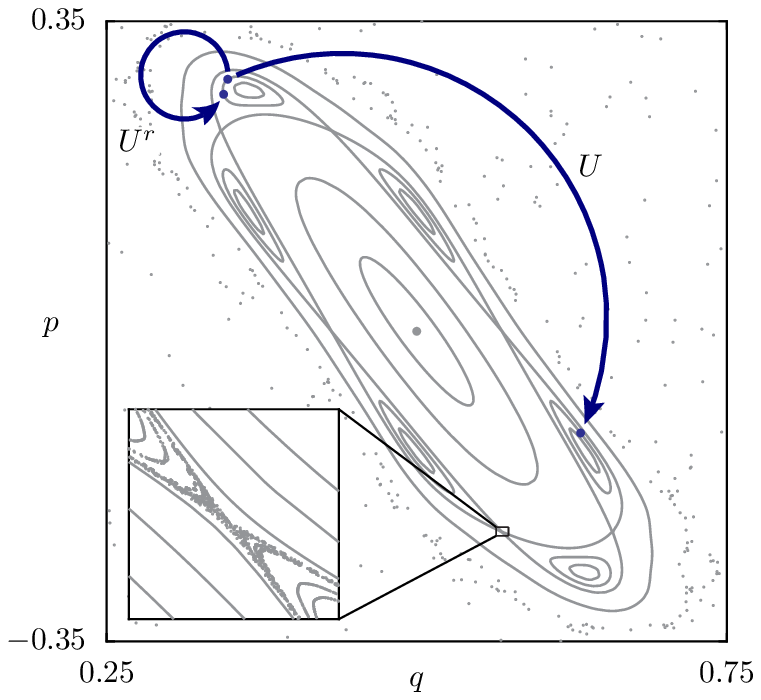}{(color online) Phase space of the standard map,
Eq.~\eqref{eq:SMap}, at $\K=3.4$ with a chaotic orbit (gray dots), regular tori (gray lines)
including the dominant $6$:$2$ resonance chain and a chaotic layer (inset).
The arrows indicate one iteration step when applying the maps $U$ and $U^r$.}{fig:SMap}
According to the KAM theorem \KAMCitations,
the regular region is composed of invariant 1D tori,
along which the iterated points rotate with a sufficiently irrational frequency $\omega$.
Following from the Poincar{\'e}--Birkhoff theorem \cite{Poi1912,Bir1913},
these irrational tori are interspersed by nonlinear $r$:$s$ resonance chains \cite{LicLie1992,Chi1979} with rational frequencies
\begin{align}
  \Ors=2\pi \frac{s}{r}.
\end{align}
E.g. for $\K=3.4$, the dominant $r$:$s=6$:$2$ resonance has
$r=6$ resonance regions, see Fig.~\ref{fig:SMap}.
Here $s$ denotes the number of resonance regions that are surpassed in one
iteration step of $U$.
Thus after $r$ periods of the external driving one has $s$ rotations around
the elliptic fixed point $(\qfix,\pfix)$.
Note that this $6$:$2$ resonance is composed of $s=2$ 
disconnected groups of $\tfrac{r}{s}=3$ resonance regions.
As the rational numbers are dense within the real numbers, there are infinitely
many nonlinear resonance chains within the regular region, where the dominant
one typically has the lowest order $r$.
Each resonance chain is surrounded by a thin chaotic layer, see the inset in Fig.~\ref{fig:SMap}.
It is important to note that applying the $r$-times iterated map $U^r$ gives
the same phase-space structure as $U$, however each resonance region
is mapped onto itself, see Fig.~\ref{fig:SMap}.

\section{Iterative canonical transformation method with a resonance}
  \label{sec:ICTM}
In this section we demonstrate how a regular phase-space region of a mixed
system and one considered nonlinear resonance chain can be approximated by an
integrable Hamiltonian $\Hrs(q,p)$.
More specifically, $\Hrs(q,p)$ is constructed such that the final point
of a time-evolution over the time span $\Delta t=r$ is close to $U^r(q,p)$,
if the initial point $(q,p)$ is chosen from the regular region.
The reason for using the resonance order $r$ as the time span $\Delta t$
instead of considering $\Delta t=1$ is indicated
in Fig.~\ref{fig:SMap}:\ Here the $r$ resonance regions
are connected by the dynamics of $U$,
a property that cannot be modeled by a time-independent integrable approximation.
We consider the $r$-fold map $U^r$ instead, where each resonance region is mapped onto itself.

In order to find $\Hrs(q,p)$, we generalize the iterative
canonical transformation method of Ref.~\cite{LoeLoeBaeKet2013}
to include the considered nonlinear resonance chain.
The iterative canonical transformation method is based on the idea,
that the tori of the regular region and their dynamics can be decomposed into the properties
(i) action and frequency as well as (ii) shape.
Accordingly, an integrable approximation is constructed in two steps:
(i) Find an integrable approximation with
matching frequencies and actions.
(ii) Transform this integrable approximation
to match the shape of the tori in phase space using iterative canonical transformations.

To include a resonance chain into the integrable approximation,
step (i) is extended to normal-form Hamiltonians
\NormalFormCitations,
as discussed in Sec.~\ref{sec:ActionFrequencyApproximation}.
This is followed by a presentation of step (ii) in Sec.~\ref{sec:ShapeApproximation}.
The specific implementation of the iterative canonical transformation method with a resonance
for the standard map is demonstrated in Sec.~\ref{sec:ICTMStd}.

\newcommand{\taulr}[1]{{#1}^\tau _{\ell r}}
\newcommand{\taulrn}[2]{{#1}^{\tau,{#2}} _{\ell r}}
\subsection{Action and frequency approximation}
  \label{sec:ActionFrequencyApproximation}

We now describe the first step of constructing an integrable
approximation to a regular phase-space region of $U^r$ and the considered nonlinear
resonance chain. This step requires to extract information about the actions
and frequencies of motion along tori in the regular phase-space region. This
information is then condensed into an integrable approximation.

\subsubsection{Extracting actions and frequencies of $U^r$}
\label{sec:ExtractActionsFrequencies}

In order to compute actions and frequencies of $U^r$, we compute the orbit
\begin{align}
 \label{eq:UOrbit}
 \taulr{\xU} &= (U^r)^\ell(\qU^\tau_0, \pU^\tau_0),
\end{align} with initial conditions $(\qU^\tau_0, \pU^\tau_0)$ for $\ell=1,...,\Nitersteps$ iterations of $U^r$.
These orbits lie on a set of tori labeled by $\tau$ in the regular region, including tori of the resonance regions.
Their action $\actU_\tau$ can be evaluated according to the general formula
\begin{align}
  \label{eq:ActionDef}
  \act = \frac{1}{2\pi}\oint_{\text{torus}} p\: \diff q.
\end{align}

Their frequency $\omegaU_\tau=\hat{\omega}/r\in[-\tfrac{\pi}{r},\tfrac{\pi}{r}[$ can be determined 
from the frequency $\hat{\omega}$ of the orbit $\hat{\x}^\tau_\ell=\taulr{\xU}$. 
Thus the orbit $\taulr{\xU}$ is described by the Fourier series 
\mbox{$\taulr{\xU}= \sum_{k\in\mathbb{Z}}\mathbf{c}^\tau_k \exp(i\omegaU_\tau \ell rk)$}.
Note that this definition of $\omegaU_\tau$
based on $U^r$ is equivalent to the definition of Ref.~\cite{LoeBaeKetSch2010}
where the frequencies of $U$ are shifted by $\Ors$ into the
co-rotating frame of the $r$:$s$ resonance.
Finally this leads to the dataset of actions and frequencies
\begin{align}
 (\actU_\tau,\omegaU_\tau)
\end{align} of the regular region of $U^r$.
Note that all quantities related to $U^r$ are
marked by an overbar to clearly distinguish them from
those quantities related to the integrable approximation.

\subsubsection{Integrable approximation}
\label{sec:NormalFormConstruction}

Based on the determined actions and frequencies
we now introduce an integrable approximation.
Following the idea of normal forms \NormalFormCitations,
we choose as an ansatz the Hamiltonian
\begin{align}
  \label{eq:HrsActAng}
  \HrsActAng(\unpang, \unpact) = \Hnod(\unpact) +\V(\unpact)\cos(r\unpang),
\end{align} where $r$ is the order of the resonance.
The phase space of this Hamiltonian consists of three integrable parts,
see Fig.~\ref{fig:area_map_Hreg}(a),
which correspond to the regular region of $U^r$ with the considered resonance chain,
see Fig.~\ref{fig:area_map_Hreg}(b).
The ansatz for $\HrsActAng(\unpang,\unpact)$ contains two arbitrary functions $\Hnod(\unpact)$ and $\V(\unpact)$.
They need to be determined according to the following criterion:
For every torus of the map $U^r$
with action $\actU_\tau$ and frequency $\omegaU_\tau$, there should (i) exist a torus of $\HrsActAng(\unpang,\unpact)$
with the same action $\act=\actU_\tau$ having (ii) a similar frequency
$\omega(\act=\actU_\tau) \approx \omegaU_\tau$. Here $\omega(\act)$ is the frequency function induced by the Hamiltonian $\HrsActAng(\unpang,\unpact)$ in the corresponding parts of phase space.

\insertfigure{}{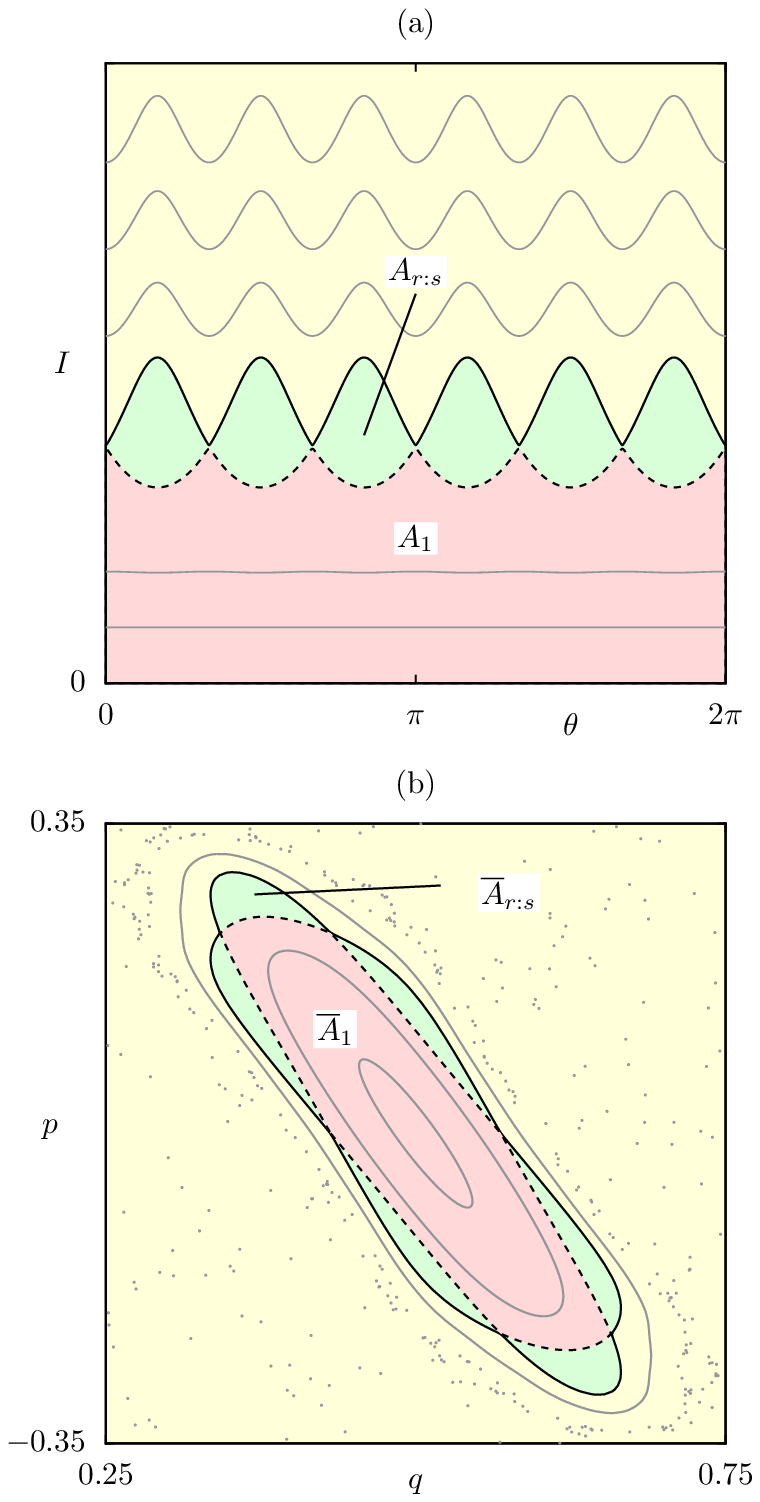}{(color online)
       (a) Phase space of the normal-form Hamiltonian
       $\HrsActAng(\unpang, \unpact)$, Eq.~\eqref{eq:HrsActAng}, with
       the areas $\Ars$ of the resonance regions and
       the area $\ArsInner$ below the resonance regions.
       (b) The corresponding areas $\ArsU$ and $\ArsInnerU$ for the standard map,
       Eq.~\eqref{eq:SMap}, at $\K=3.4$.}{fig:area_map_Hreg}
To achieve (i), $\Hnod(\unpact)$ and $\V(\unpact)$ are chosen such that
the total area $\Ars$ of the resonance regions and
the area $\ArsInner$ below the resonance region
agree with the corresponding areas of $U^r$, see Fig.~\ref{fig:area_map_Hreg},
\begin{subequations}
  \label{eq:AreaCondition}
\begin{align}
 \Ars &\approx \ArsU, \\
 \ArsInner & \approx \ArsInnerU.
\end{align}
\end{subequations}
To achieve (ii), we further choose $\Hnod(\unpact)$ and $\V(\unpact)$ such that the distance of corresponding frequencies in $U^r$ and $\HrsActAng$,
\begin{align}
  \label{eq:FrequencyFit}
  \sum_{\tau} \left| \omegaU_{\tau} - \omega(\actU_{\tau})\right|^{2},
\end{align} is minimized.
An explicit determination of $\Hnod(\unpact)$ and $\V(\unpact)$ from these conditions 
in terms of a series expansion is demonstrated 
in Sec.~\ref{sec:ActionFrequencyApproximationStd}
for the example of the standard map.

\subsection{Shape approximation}
  \label{sec:ShapeApproximation}

We now show how the second step of the iterative canonical transformation method is implemented. 
For this the normal-form Hamiltonian $\HrsActAng(\unpang, \unpact)$ with adapted frequencies is
transformed to the phase-space coordinates $(q,p)$ such that its time-evolution over the time span $\Delta t=r$
closely agrees with $U^r$ in the regular phase-space region.
To achieve this the transformed tori of the integrable approximation should
match the shape of the corresponding tori in the regular phase-space region of $U^r$ including the considered
nonlinear resonance chain.
For this we adapt the iterative canonical transformation method \cite{LoeLoeBaeKet2013}
to the case of an additional resonance chain:
In Sec.~\ref{sec:InitialIntegrableApproximation} we explain how an initial canonical
transformation is used to find an initial integrable approximation which roughly
resembles the regular phase-space region of the mixed system including the considered resonance
chain.
In Sec.~\ref{sec:FamilyOfCanonicalTransformations} we introduce a family
of canonical transformations.
In Sec.~\ref{sec:IterativeImprovement} we explain how iterative application
of canonical transformations gives an improved integrable approximation.

\subsubsection{Initial integrable approximation}
  \label{sec:InitialIntegrableApproximation}

In order to transform the normal-form Hamiltonian $\HrsActAng(\unpang,\unpact)$
to the phase-space coordinates
$(q,p)$ of the regular phase-space region of $U^r$, we apply an initial
canonical transformation
\begin{align}
  \label{eq:T0}
  T_0: (\unpang,\unpact) \mapsto (q,p).
\end{align}
This initial canonical transformation $T_0$ should map the tori of
$\HrsActAng(\unpang,\unpact)$ to the neighborhood of the corresponding tori of $U^r$.
In particular the torus with action $\act=0$ should be mapped
onto the fixed point $(\qfix,\pfix)$ of $U^r$.
This gives the initial integrable
approximation
\begin{align}
  \label{eq:Hrs0}
  \Hrs^{0}(q,p) = \HrsActAng\left[T_{0}^{-1}(q,p)\right].
\end{align}
It is convenient to choose $T_0$ in a simple closed form,
see Sec.~\ref{sec:ShapeApproximationStd} for an example.

\subsubsection{Family of canonical transformations}
  \label{sec:FamilyOfCanonicalTransformations}

In the following we improve the agreement between the tori of the initial integrable
approximation $\Hrs^0(q,p)$ and those of the regular phase-space region of $U^r$.
To this end we introduce a family of type two generating functions
\begin{align}
  \label{eq:GeneratingFunction}
  F^{\afam}(q, p') = qp' + \sum_{\nu=1}^\Nparams a_\nu G_\nu(q,p'),
\end{align}
defined by a choice of $\Nparams$ parameters $\afam =(a_1,a_2,...,a_\Nparams) \in \mathbb{R}^\Nparams$ and a choice of
independent functions $G_\nu$. The corresponding canonical transformation
\begin{align}
  \label{eq:CanonicalTransformationQP}
  T^{\afam}:(q,p)\mapsto(q',p')
\end{align}
is implicitly defined by the equations
\begin{subequations}
  \label{eq:CanonicalTransformation}
  \begin{align}
  q' =& \dfrac{\partial F^{\mathbf{a}}}{\partial p'}(q, p') = q + \sum_{\nu=1}^\Nparams
a_\nu \dfrac{\partial G_\nu(q,p')}{\partial p'}, \\
  p =& \dfrac{\partial F^{\mathbf{a}}}{\partial q}(q, p') = p' + \sum_{\nu=1}^\Nparams
a_\nu \dfrac{\partial G_\nu(q,p')}{\partial q},
  \end{align}
that need to be solved for $(q',p')$. For sufficiently small $\afam$
and bounded $C^2$ functions $G_\nu$ this solution globally exists 
according to Hadamard's global inverse function theorem \cite{KraPar2002} 
and represents a near-identity transformation.
\end{subequations}

\subsubsection{Iterative improvement}
  \label{sec:IterativeImprovement}

We now use a family of canonical transformations $T^\afam$ to improve the
agreement between the initial integrable approximation $\Hrs^0(q,p)$, Eq.~\eqref{eq:Hrs0},
and the regular phase-space region of $U^r$.
From a theoretical point of view it is tempting to find a canonical
transformation $T^{\afam}$ leading to a new Hamiltonian which shows maximal agreement with the
regular phase-space region of $U^r$.
However, finding this transformation, e.\,g., by making an ansatz for $T^\afam$ 
using Fourier basis functions $G_\nu$ in Eqs.~\eqref{eq:GeneratingFunction} 
and \eqref{eq:CanonicalTransformation} with an infinite set of coefficients, 
is practically impossible.
Therefore, we fix the number $\Nparams$ of coefficients in our ansatz 
for the family of canonical transformations $T^{\afam}$. 
Subsequently, we use members from this family
to iteratively improve the agreement between the integrable
approximation and the regular phase-space region of $U^r$.
This gives a sequence of canonical transformations
\begin{align}
  T_{n}:(q,p)\mapsto(q',p'), \quad n=1,2,...,N,
\end{align}
with $T_{n}\in \{T^{\afam}\}$ such that the $n$th integrable approximation
\begin{align}
  \label{eq:Hrsn}
  \Hrs^{n}(q,p) = \Hrs^{0}\left[ T_{1}^{-1}\circ \dots \circ
                  T_{n}^{-1}(q,p)\right],
\end{align}
agrees more and more with the regular phase-space region
of $U^r$ when $n$ is increased.

For this, each canonical transformation has to minimize the distance of points with corresponding action--angle coordinates,
$(\ang,\act)=(\angU,\actU)$, in $\Hrs^n$ and $U^r$, respectively.
To achieve this
(i) we explain how to obtain the corresponding sample points and
(ii) we set up a cost function to minimize their distance.

(i) Using the orbit of Eq.~\eqref{eq:UOrbit}, we obtain the sample points $\taulr{\xU}$ of $U^r$,
which correspond to action $\actU_\tau$ and angles
\begin{align} \label{eq:SampleAnglesU}
 \taulr{\angU} &= \omegaU_\tau \ell r.
\end{align}
For the integrable approximation we first define the corresponding sample points of $\HrsActAng(\unpang,\unpact)$,
\begin{subequations}
  \label{eq:PointsThetaI}
\begin{align}\label{}
 \taulr{\unpang} &= \unpang(\ang,\act)\Big\vert_{(\ang,\act)=(\taulr{\angU}, \actU_\tau)},\\
 \taulr{\unpact} &= \unpact(\ang,\act)\Big\vert_{(\ang,\act)=(\taulr{\angU}, \actU_\tau)}.
\end{align}
\end{subequations}
Here $(\ang,\act)$ denote the action--angle coordinates 
of $\HrsActAng(\unpang,\unpact)$ which exist, 
as $\HrsActAng(\unpang,\unpact)$ is locally integrable.
If the used transformation $\unpang(\ang,\act),\unpact(\ang,\act)$
% why is a backslash space needed here?
is known explicitly, as e.\,g.\ for the pendulum Hamiltonian \cite{LicLie1992},
an evaluation of Eqs.~\eqref{eq:PointsThetaI} is straightforward.
If this transformation is not known explicitly, which is typically the case,
we construct $(\taulr{\unpang},\taulr{\unpact})$
using the time evolution with $\HrsActAng(\unpang,\unpact)$.
More specifically, we choose $(\unpang^\tau _0, \unpact^\tau _0)$ 
to be the point on the torus of action $\actU_\tau$ which 
is closest to $T_0^{-1}(\xU^\tau _0)$.
We then obtain the points $(\taulr{\unpang}, \taulr{\unpact})$ 
from an evolution with $\HrsActAng(\unpang,\unpact)$ up to the time $t=\ell r f$.
Here, the factor $f=\omegaU_\tau/\omega(\actU_\tau)$ is of order $1$ and ensures 
that the angle $\ang=\omega(\actU_\tau)t=\omegaU_\tau \ell r$ agrees with 
the corresponding angle $\taulr{\angU}$ of $U^r$, Eq.~\eqref{eq:SampleAnglesU}.
Finally this gives the sample points of the $n$th integrable approximation $\Hrs^n(q,p)$,
\begin{align}
   \taulrn{\x}{n} &= T_n\circ\dots\circ T_1\circ T_0 (\taulr{\unpang},\taulr{\unpact}),
\end{align} which correspond to the sample points $\taulr{\xU}$ of $U^r$.

(ii)
To minimize the distance between $\xU^\tau _{\ell r}$ and $\x^{\tau,n}_{\ell r}$ in the $(n+1)$st iteration step,
we apply the canonical transformation $T^\afam$ and minimize the cost function
\begin{align}
  \label{eq:CostFunction}
  L(\afam) = \frac{1}{\Nsamplepoints} \sum_\tau \sum_\ell
                   \left[ \taulr{\xU} -
T^{\afam}\left(\taulrn{\x}{n}\right)\right]^2.
\end{align}
Here $\Nsamplepoints$ is the total number of sample points.
Since $\afam=\bfzero$ gives the identity transformation according to
Eq.~\eqref{eq:CanonicalTransformation},
$L(\bfzero)$ measures the quality of $\Hrs^n$.
Thus any choice of $\afam$ with $L(\afam)<L(\bfzero)$ improves $\Hrs^n$.

Furthermore, following the strategy of Ref.~\cite{LoeLoeBaeKet2013}, we determine an
optimal parameter $\afam$. To this end we exploit that $\Hrs^{0}(q,p)$ agrees
well with the approximated phase-space region already, such that the optimal
transformation should be close to the identity transformation,
i.\,e. the sought-for parameter $\afam$ is small,
\begin{align}
  |\afam|\ll1.
\end{align}
This allows for solving Eq.~\eqref{eq:CanonicalTransformation} to linear
order, giving a quadratic approximation to the cost function
$L(\afam)$ \cite{LoeLoeBaeKet2013}. From this a good estimate of the optimal parameter $\afam^{*}$ close
to the minimum of $L(\afam)$ is determined. For this parameter $\afam^{*}$
one solves the canonical transformations~\eqref{eq:CanonicalTransformation} numerically using Newton's method.
If for this parameter $\afam^{*}$ Eq.~\eqref{eq:CanonicalTransformation} 
is not invertible on the relevant domain of phase space,
we replace $T^{\afam^*}$ by $T^{\eta\afam^*}$ using a damping factor $\eta\ll 1$.
This is possible as $L(\afam^*)<L(\eta\afam^*)<L(\bfzero)$, 
but requires to increase the number $N$ of iteration steps.

\section{Application to the standard map}
  \label{sec:ICTMStd}

In this section we describe how the iterative canonical transformation method
with a resonance is implemented for the central regular
phase-space region of the standard map.
We first consider this map, Eq.~\eqref{eq:SMap}, for $\K=3.4$,
where it has a nonlinear $6$:$2$ resonance, see Fig.~\ref{fig:SMap}.

\subsection{Action and frequency approximation}
  \label{sec:ActionFrequencyApproximationStd}

\subsubsection{Extracting actions and frequencies of $U^r$}
\label{sec:ExtractActionsFrequenciesStd}

According to Sec.~\ref{sec:ExtractActionsFrequencies} we start by determining
the actions and frequencies
$(\actU_{\tau},\omegaU_{\tau})$ from the regular phase-space region of the map $U^r$.
For a rough scan of the regular region, we consider a set of points
on a line at $\pU_0=\pfix$ with $\qU_0(\crvparam)=\qfix+\crvparam$
using equidistant parameter values $\crvparam\in]0,0.0931]$.
To each of these points we apply the map $U^r$ to obtain an orbit
and determine its frequency $\omegaU(\crvparam)$ \cite{LasFroCel1992,BarBazGioScaTod1996}.
Since frequencies change in a non-smooth way across the infinitely many resonances of the regular region,
we focus on so-called noble tori, which are furthest away from these resonances.
We determine a set of $\Ntori=80$ target frequencies $\omegaU_\tau$ 
from the range of frequencies $\omegaU(\crvparam)$
as described in Appendix~\ref{sec:NobleFrequencies}.
For each target frequency $\omegaU_\tau$ 
we solve $\omegaU(\crvparam_\tau)=\omegaU_\tau$ for $\crvparam_\tau$ numerically. 
\insertfigure{}{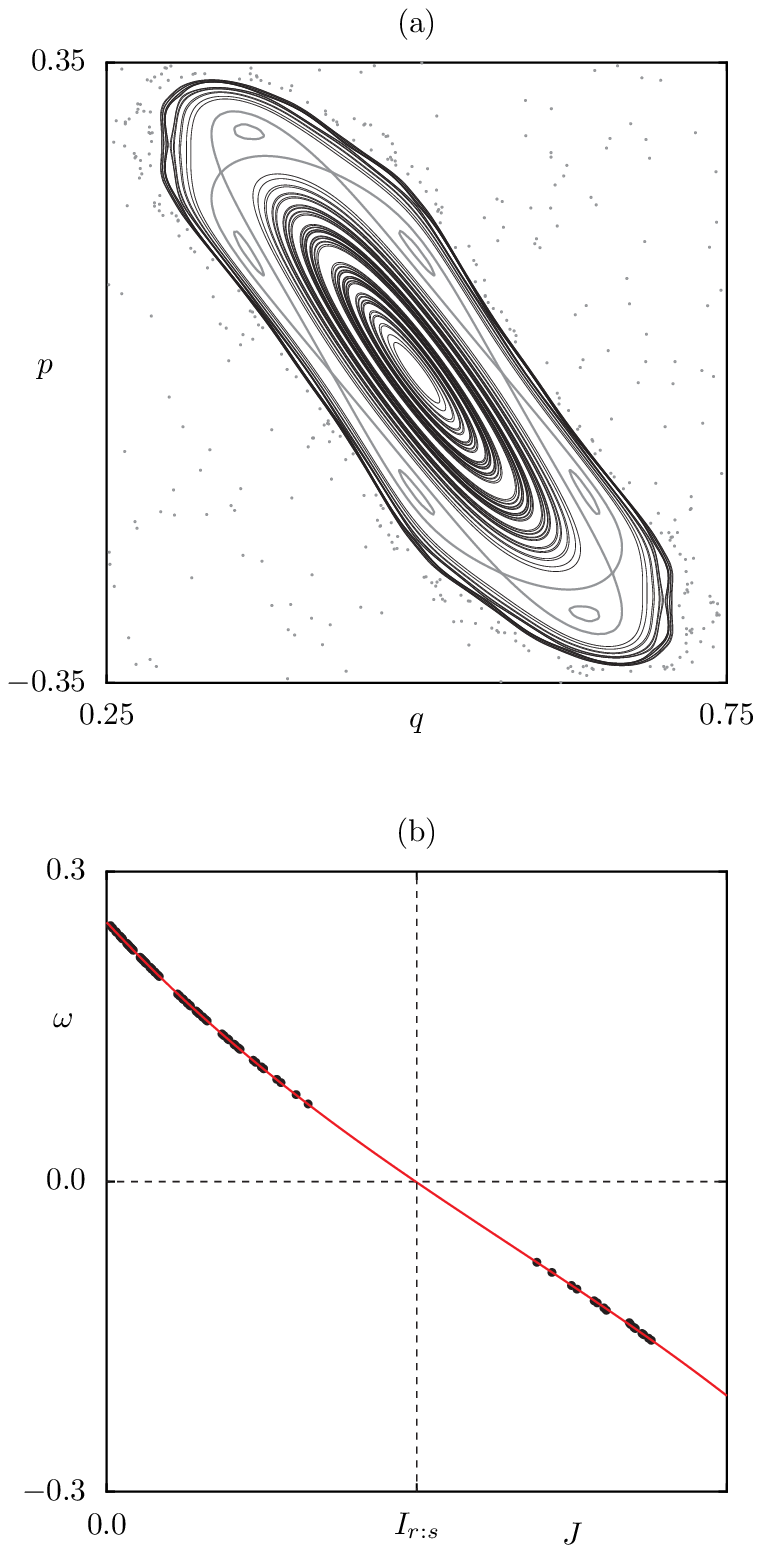}{%
       (color online) (a) Phase space of the standard map,
       Eq.~\eqref{eq:SMap}, at $\K=3.4$ (gray lines and dots) with regular orbits $\taulr{\xU}$ (black)
       on noble tori.
       (b) Frequencies $\omegaU_\tau$ of these orbits (black dots)
       and the fitted function $\Hnod'(\act)$ (red line).%
       }{fig:tori_and_frequencies1}
\insertfigure{}{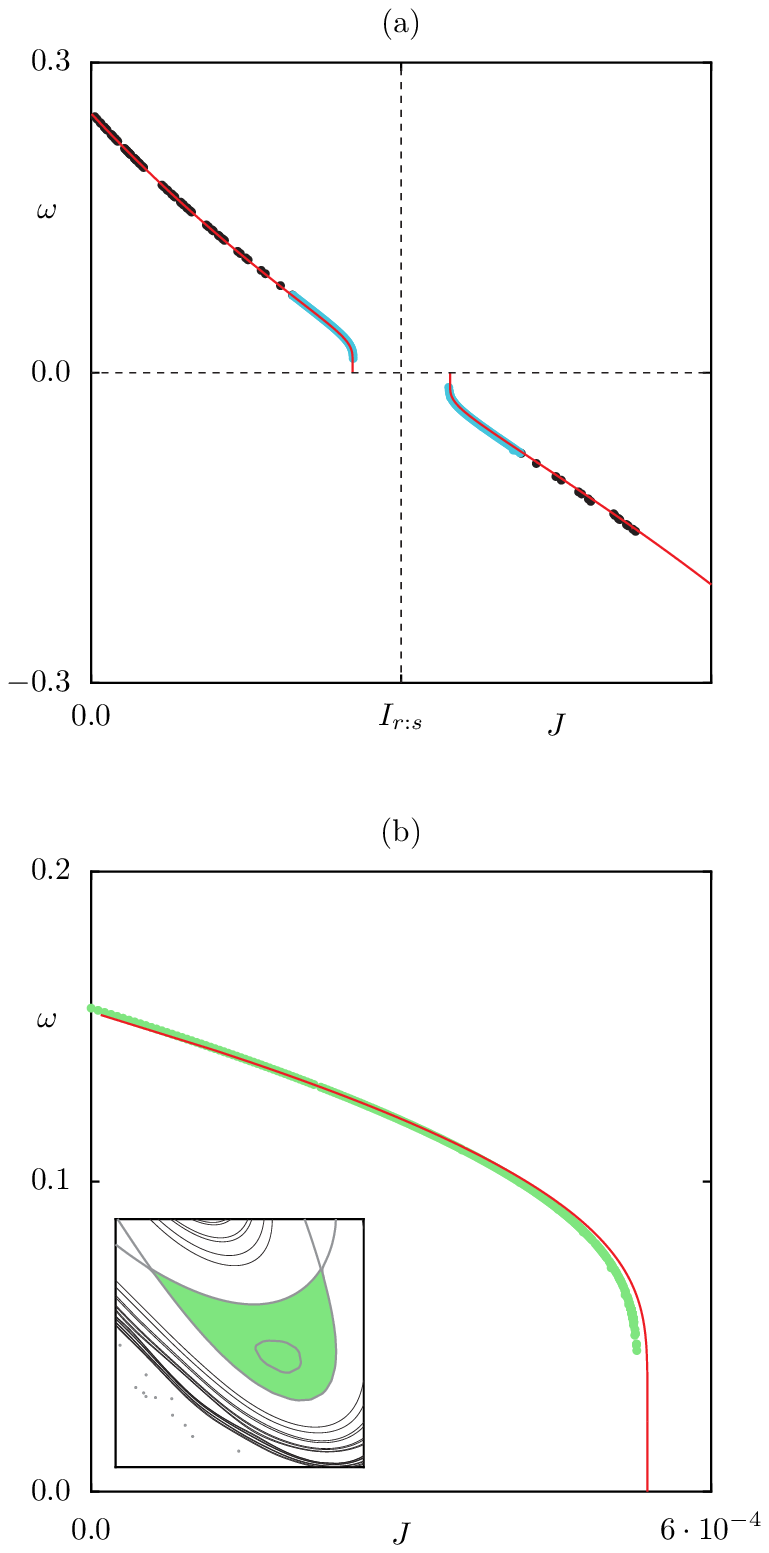}{%
       (color online) Comparison of the frequency function $\omega(\act)$
       of the determined integrable approximation $\HrsActAng(\unpang,\unpact)$ (red lines)
       to frequencies of $U^r$ (dots): (a) the frequencies $\omegaU_\tau$ (black dots), 
       frequencies close to the resonance region (light blue dots) 
       and (b) frequencies inside the resonance region (green dots).}{fig:tori_and_frequencies2}

This gives a set of initial conditions $(\qU^\tau_0, \pU^\tau_0)=(\qU_0(\crvparam_\tau), \pU_0)$
on noble tori $\tau$.
From these initial conditions we compute the orbits $\xU^\tau _{\ell r}$,
Eq.~\eqref{eq:UOrbit}, using $\Nitersteps=10^4$ iterations of the map $U^r$,
resulting in the black tori shown in Fig.~\ref{fig:tori_and_frequencies1}(a).
We compute their action $\actU_\tau$ according to Eq.~\eqref{eq:ActionDef}.
This gives the dataset of actions and frequencies
$(\actU_\tau, \omegaU_{\tau})$ which is depicted
by the black dots in Fig.~\ref{fig:tori_and_frequencies1}(b).
Note that a similar procedure could be applied to the tori inside the considered resonance chain.
However, for convenience we do not use those tori which will turn out to be sufficient.

\subsubsection{Integrable approximation}
\label{sec:NormalFormConstructionStd}

\insertbigfigure{}{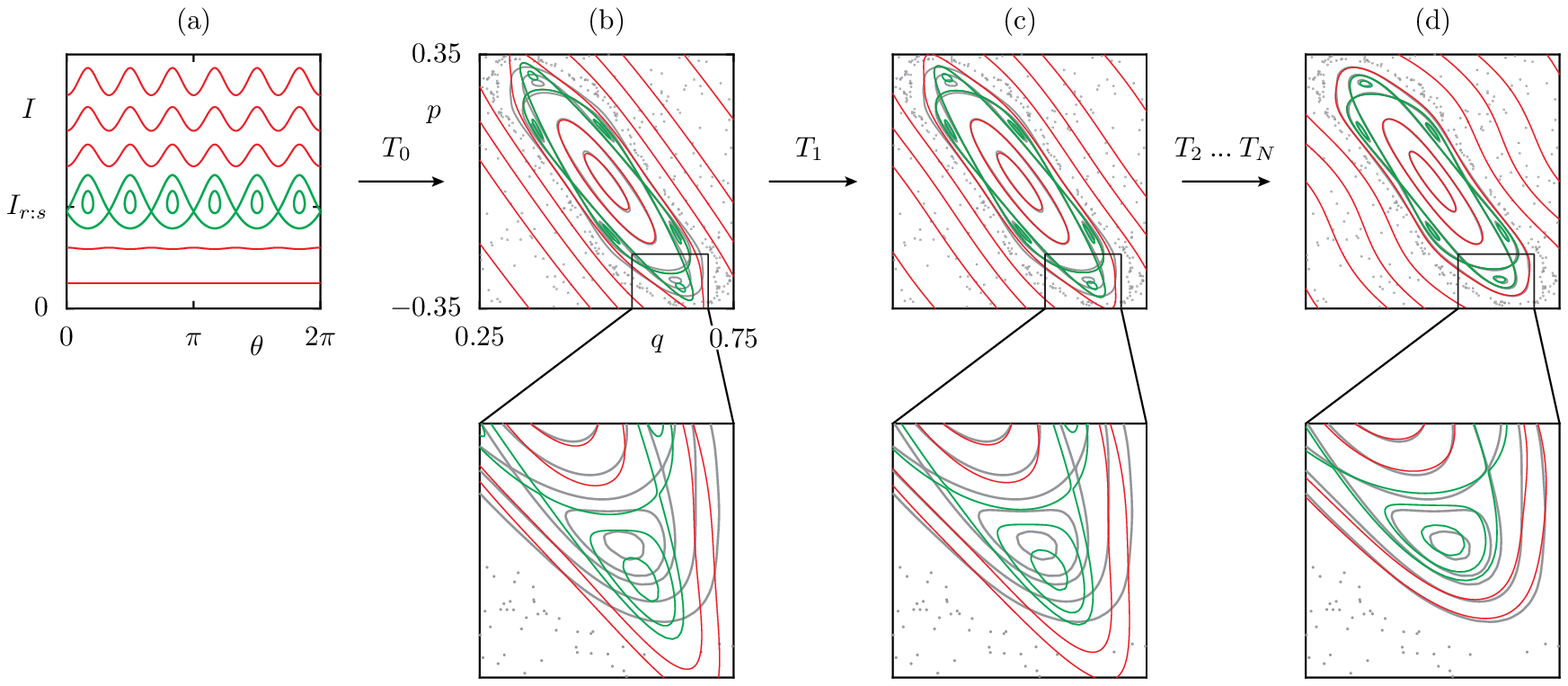}{%
       (color online) (a) Phase space of the
       normal-form Hamiltonian $\HrsActAng(\unpang, \unpact)$,
       Eq.~\eqref{eq:HrsActAng}, with $\Hnod(\unpact)$ and $\V(\unpact)$ 
       as determined in Sec.~\ref{sec:ActionFrequencyApproximationStd} (thin colored lines).
       (b--d) Phase space of the standard map,
       Eq.~\eqref{eq:SMap}, at $\K=3.4$ (light gray lines and dots)
       and tori (thin colored lines)
       of the integrable approximations $\Hrs^n(q,p)$ obtained from the transformation $T_n\circ\hdots\circ T_1\circ T_0$ 
       (b) $\Hrs^0(q,p)$,
       (c) $\Hrs^1(q,p)$, and 
       (d) $\Hrs^N(q,p)$, $N=15$. The magnifications show the improvement of the integrable approximations.}{fig:Hreg0_HregN}
As explained in Sec.~\ref{sec:ActionFrequencyApproximation} we now require a
normal-form Hamiltonian
which matches the corresponding actions and frequencies of the standard map $U^r$
by satisfying the area conditions~\eqref{eq:AreaCondition} and
minimizing Eq.~\eqref{eq:FrequencyFit}.
For this normal-form Hamiltonian $\HrsActAng(\unpang, \unpact)$, Eq.~\eqref{eq:HrsActAng},
we use
\begin{align}
 \Hnod(\unpact) &= \frac{(\unpact-\Irs)^2}{2\Mrs} + \sum_{k=3}^{\Nmaxorder}h_k (\unpact-\Irs)^k,\label{eq:DispersionAnsatzStd}
\end{align} and the lowest order ansatz for a resonance chain encircling a fixed point
\cite{LoeBaeKetSch2010,LebMou1999,DeuMouSch2013},
\begin{align}
 \V(\unpact) &= 2\Vrs \left(\frac{\unpact}{\Irs}\right)^{r/2}.\label{eq:VAnsatzStd}
\end{align}
In order to determine the unknown parameters $\{\Irs,\Mrs,\Vrs,h_k\}$ we analyze $\HrsActAng(\unpang,\unpact)$ first close to the resonance and secondly far away from the resonance.

Close to the resonance, the leading order expansion of $\HrsActAng(\unpang,\unpact)$ around $\Irs$ is the pendulum Hamiltonian 
\cite{Alm88,BroSchUll2001,BroSchUll2002}
\begin{align}
 \Hpend(\unpang,\unpact) &= \frac{(\unpact-\Irs)^2}{2\Mrs}+2\Vrs\cos(r\unpang).
\end{align}
Here $\Irs$ gives the location of the resonance, 
while $\Mrs$ and $\Vrs$ control the size $\Ars$ of the resonance and the frequency at the center of the resonance region.
We compute these parameters according to \cite{EltSch2005}
\newcommand{\arccosTrM}{\arccos\left(\tfrac{1}{2}\text{Tr}\,\monodromy\right)}
\begin{subequations}\label{eq:IrsMrsVrs}
\begin{align}
 \Irs &= \frac{1}{2\pi}(\ArsInnerU+\tfrac{1}{2}\ArsU),\\
 \Mrs &= \frac{\signdisp r^2}{16}\ArsU\arccosTrM^{-1},\\
 \Vrs &= \frac{\signdisp}{32 r^2}\ArsU\arccosTrM.
\end{align}
\end{subequations} This accounts for condition~\eqref{eq:AreaCondition}
by matching the areas $\ArsU$ and $\ArsInnerU$ of $U^r$, see Fig.~\ref{fig:area_map_Hreg}.
Furthermore, the frequency at the center of the resonance region
enters via the monodromy matrix $\monodromy$.
Note that these parameters contain the essential information 
on action and frequency within the resonance regions.
Finally, we find for the sign $\signdisp=-1$, because the frequencies decrease with increasing action, see Fig.~\ref{fig:tori_and_frequencies1}(b).

We now determine the parameters $\{h_k\}$ which describe the frequency behavior far away from the resonance regions.
There the frequency function $\omega(\act)$ is approximately described by
\begin{align}\label{eq:FrequencyApproximation}
\omega(\act)  &\approx\Hnod'(\act) = \frac{\act-\Irs}{\Mrs} + \sum_{k=3}^{\Nmaxorder}k h_k (\act-\Irs)^{k-1},
\end{align} which neglects the resonance as a perturbation.
In this approximation, Eq.~\eqref{eq:FrequencyFit} becomes
\begin{align}
 \sum_\tau\left|\omegaU_\tau-\Hnod'(\actU_\tau)\right|^2 \label{eq:FrequencyFitApproximation},
\end{align} which we minimize to determine $\{h_k\}$.
For $\Nmaxorder=4$ we obtain a satisfactory agreement
between the dataset $(\actU_\tau,\omegaU_\tau)$ and the approximate frequency function $\Hnod'(\act)$,
see Fig.~\ref{fig:tori_and_frequencies1}(b).
Note that this comparison is meaningful only far from the resonance, where the approximation~\eqref{eq:FrequencyApproximation} is justified.

The determined parameters give the resulting Hamiltonian $\HrsActAng(\unpang,\unpact)$, see Fig.~\ref{fig:Hreg0_HregN}(a).
For a global comparison, we perform a numerical evaluation of the exact frequency function $\omega(\act)$ of $\HrsActAng(\unpang,\unpact)$.
We obtain a good agreement with a mean error of $\Delta\omega=0.0002$ for the dataset $(\actU_\tau,\omegaU_\tau)$ and
also near the resonance (light blue dots in Fig.~\ref{fig:tori_and_frequencies2}(a)) we have $\Delta\omega<0.001$.
Moreover, even inside the resonance regions where no data of tori has been used for the optimization,
but only the parameters of Eqs.~\eqref{eq:IrsMrsVrs}, the frequency is well approximated, see Fig.~\ref{fig:tori_and_frequencies2}(b).

\subsection{Shape approximation}
  \label{sec:ShapeApproximationStd}

We proceed by mapping the integrable approximation 
obtained in the previous section to the phase space of $U^r$.
As a first step, we choose the initial canonical transformation
\begin{align}
  \label{eq:T0_std}
  T_{0}:\quad	\left(\begin{array}{c} \unpang \\ \unpact \end{array}\right)
		\mapsto
		\left(\begin{array}{c} \qfix \\ \pfix \end{array}\right)
        + \mathcal{R}
        \left(\begin{array}{c} \sqrt{2\act}\cos(\unpang) \\ -\sqrt{2\act}\sin(\unpang) \end{array} \right)
\end{align}
with
\begin{align}
  \label{eq:R_std}
  \mathcal{R} = \left(\begin{array}{cc}
						1 & 1/2 \\
						0 & 1
					  \end{array}
                \right)
				\left(\begin{array}{cc}
						1/\sqrt{\sigma} & 0 \\
						0 & \sqrt{\sigma}
					  \end{array}
                \right).
\end{align}
The parameter $\sigma$ of $T_0$ is chosen such that the 
hyperbolic periodic points of the nonlinear resonance chain along the line $p=0$ agree both for
the standard map and the induced initial integrable approximation $\Hrs^{0}(q,p)$, Eq.~\eqref{eq:Hrs0}.
The specific choice for $T_0$
incorporates the symmetries of the standard map into the initial integrable
approximation.
The result for $\K=3.4$ using $\sigma=3.96851$ is depicted in
Fig.~\ref{fig:Hreg0_HregN}(b).

\insertfigure{bt}{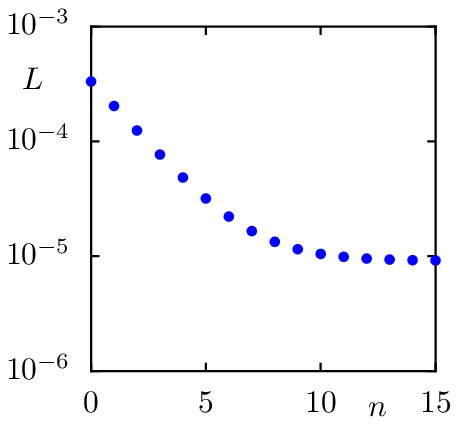}{(color online) Cost function $L$, 
  Eq.~\eqref{eq:CostFunction}, vs. iteration step $n$.}{fig:CostFunction}
To improve the initial integrable approximation we define the family of
canonical transformations $T^{\afam}$ using the Fourier ansatz for the generating function
\begin{align}
  \label{eq:GeneratingFunctionStd}
  F^{\afam}(q, p') &= qp' \\ \nonumber
                   &+ \sum_{\nu_{1}=0}^{\Nq}\sum_{\nu_{2}=0}^{\Np}
a^{+}_{\nu_{1}\nu_{2}} f_{\nu_{1}}^{+}\left(\frac{q-\qfix}{\Lq}\right)
f_{\nu_{2}}^{+}\left(\frac{p-\pfix}{\Lp}\right) \\ \nonumber
                   &+ \sum_{\nu_{1}=1}^{\Nq}\sum_{\nu_{2}=1}^{\Np}
a^{-}_{\nu_{1}\nu_{2}} f_{\nu_{1}}^{-}\left(\frac{q-\qfix}{\Lq}\right)
f_{\nu_{2}}^{-}\left(\frac{p-\pfix}{\Lp}\right),
\end{align}
with basis functions
\begin{subequations}
  \label{eq:BasisFunction}
  \begin{align}
	f_{\nu}^{+}(x) &= \cos(2\pi\nu x), \\
	f_{\nu}^{-}(x) &= \sin(2\pi\nu x).
  \end{align}
\end{subequations}
This ansatz gives canonical transformations,
Eq.~\eqref{eq:CanonicalTransformationQP}, which preserve the 
parity of the
standard map. Since shifting the generating function by a constant term is
irrelevant for the canonical transformation, we set $a_{00}^{+}=0$.
Furthermore we choose
$\Lq=\Lp=1.1$ and $\Nq=\Np=3$.

In order to set up the cost function $L(\afam)$,
Eq.~\eqref{eq:CostFunction}, we compute the sample points $\taulr{\xU}$
within the regular phase-space region of the standard map,
using Eq.~\eqref{eq:UOrbit} with $\Nitersteps=10^3$ iterations
for the same initial conditions $\xU^\tau _0$ as in Sec.~\ref{sec:ActionFrequencyApproximationStd}.
Hence, $\taulr{\xU}$ are points on noble tori of action $\actU_{\tau}$
and frequency $\omegaU_{\tau}$.
We compute the corresponding sample points $\taulrn{\x}{0}$ of
$\Hrs^0(q,p)$ by numerical integration over times $t=\ell r \omegaU_\tau/\omega(\actU_\tau)$, 
as explained in Sec.~\ref{sec:IterativeImprovement}.
For this we use initial conditions $\x^{\tau,0}_0$ on the line $p=\pfix$, $q>\qfix$,
such that the corresponding tori have action $\actU_{\tau}$.

Having defined the sample points $\taulr{\xU}$ and $\taulrn{\x}{0}$,
we now minimize the cost function $L(\afam)$,
Eq.~\eqref{eq:CostFunction}, according to the procedure described in
Sec.~\ref{sec:IterativeImprovement}, i.\,e., we iteratively determine and apply
canonical transformations $T_n$ from the family of canonical transformations defined by
Eq.~\eqref{eq:GeneratingFunctionStd}. Here, we use the damping factor
$\eta=0.25$. Applying $N=15$ steps of the iterative canonical transformation
method, we typically observe a saturation of the cost function,
see Fig.~\ref{fig:CostFunction}, giving a
sequence of improved integrable approximations $\Hrs^{n}(q,p)$ as shown in
Fig.~\ref{fig:Hreg0_HregN}.
The final integrable approximation $\Hrs=\Hrs^N$ gives a very good description of
the regular region and the $6$:$2$ resonance regions.
Even the tori inside the resonance regions
which have not yet been included in the cost function,
are well approximated.

In Fig.~\ref{fig:HregGallery} we show integrable approximations for further parameters
$\K=2.9$, $3.3$, $3.5$ of the standard map also including a case with a $10$:$3$ resonance.
Here we used the same procedure with parameters 
$\Ntori=80$, $\Nmaxorder=4$, $\Lq=\Lp=1.1$, $\Nq=\Np=3$, $\Nitersteps=10^3$, 
and damping factors $\eta=0.1$, $0.4$, $0.25$, respectively.
This demonstrates the general applicability of the presented method.

\section{Summary and outlook}
  \label{sec:Summary}

\insertbigfigure{t}{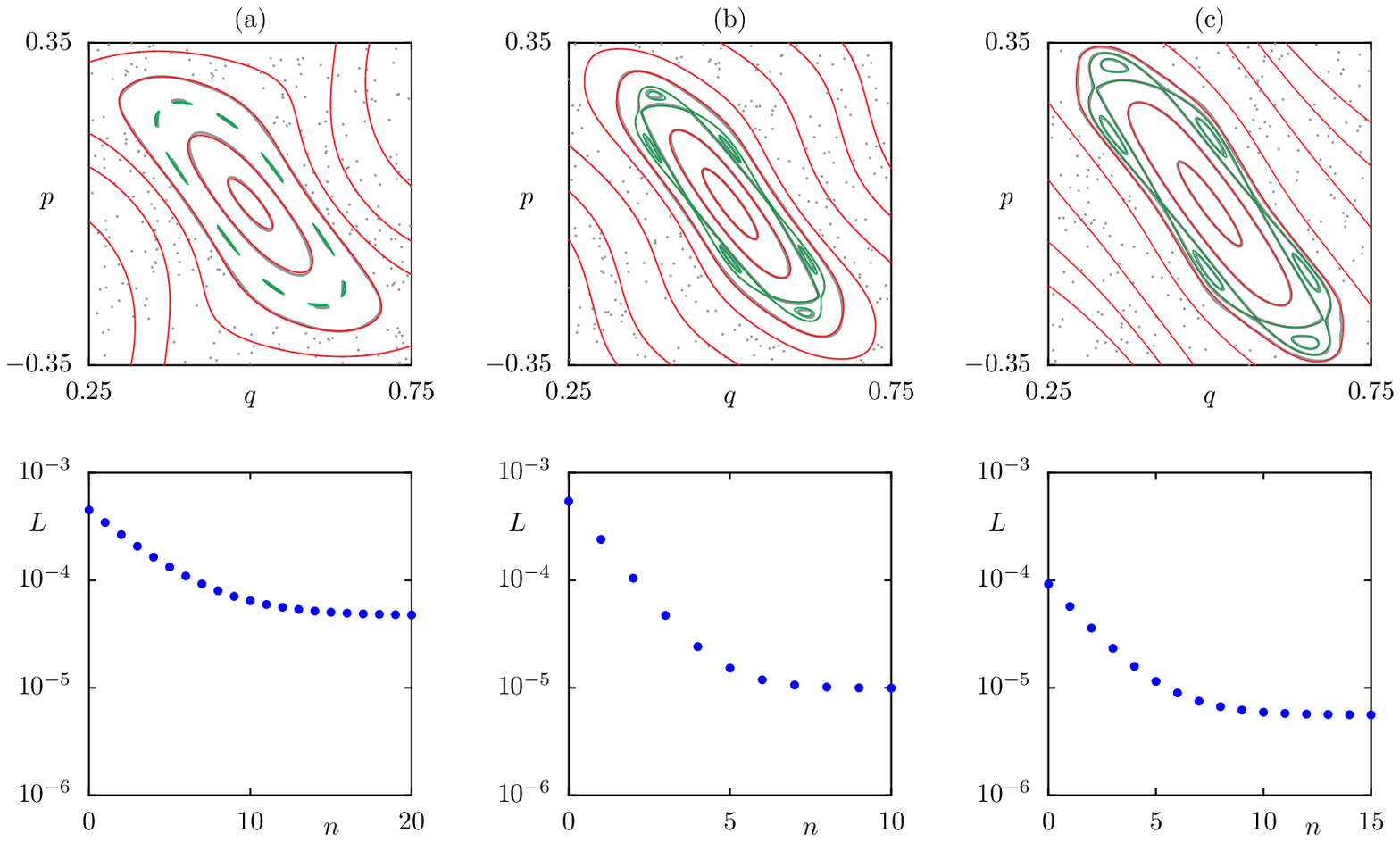}{%
       (color online) Integrable approximations for the standard map, Eq.~\eqref{eq:SMap},
       at different parameters (a) $\K=2.9$, (b) $\K=3.3$, and (c) $\K=3.5$.
       Top:
       Phase space of the standard map (light gray lines and dots)
       and tori of the integrable approximation (thin colored lines).
       Bottom:
       Cost function $L$,
       Eq.~\eqref{eq:CostFunction}, vs. iteration step $n$.%
       }{fig:HregGallery}
In this paper we  present how an integrable approximation can be
constructed to the regular phase-space region of a mixed system and one
nonlinear resonance chain.
In order to achieve this goal we combine the theory of normal-form Hamiltonians
with the iterative canonical transformation method. 
We apply this approach to the generic standard map
for various parameter values and
find an integrable approximation which closely resembles the dynamics in the
regular phase-space region including the considered resonance chain.

One possible generalization of this approach would be
to approximate multiple resonance chains.
This would require normal-form Hamiltonians with
more than one nonlinear resonance chain, which is the topic of current research \cite{PvtComm}.
Another generalization would be
the application to systems with a higher-dimensional phase space.
Here the main difficulty is to find an integrable normal-form Hamiltonian
with tori of appropriate actions and frequencies.
On the other hand, the shape approximation using the iterative canonical transformation method
should be straightforward.

\begin{acknowledgments}
We thank J\'er\'emy Le Deunff and Peter Schlagheck for stimulating discussions.
Furthermore, we acknowledge support by the Deutsche Forschungsgemeinschaft 
(Germany) within the Forschergruppe 760 \textit{Scattering Systems with 
Complex Dynamics}. 
N.\,M. acknowledges support by JSPS (Japan) grant No.\,PE 14701.
J.\,K., C.\,L., and N.\,M. contributed equally to this work.
\end{acknowledgments}

\begin{appendix}

\section{Determination of noble frequencies
           \label{sec:NobleFrequencies}}

In this appendix we describe the determination of frequencies
$\omegaU_\tau$ of noble tori $\tau$ inside the regular region.
According to the KAM theorem \KAMCitations tori persist, for which $\omegaU_\tau/(2\pi)$
is sufficiently irrational, i.e.\
satisfies a Diophantine condition.
This is for example fulfilled
for noble numbers whose continued fraction expansion is
eventually periodic with 1.
Such noble numbers are as far as possible away from rationals
in the sense that they are hardest to approximate by rationals \cite{HarWri1975}.
Thus noble tori are particularly suited
for the iterative canonical transformation method.

For convenience, we relate the frequencies $\omegaU_\tau$
to numbers $\xi_\tau\in[0,1[$ by 
\begin{align}\label{eq:FrequencyXiRelation}
 \xi_\tau &= \frac{\omegaU_\tau}{2\pi}\mod 1.
\end{align}
We now calculate $\Ntori$ noble numbers $\xi_\tau$.
This is done by first constructing
the Stern--Brocot tree \cite{Ste1858,Bro1860,GraKnuPat1994}
of rational numbers and then determining corresponding noble numbers.

1.) To build the Stern--Brocot tree in the interval $[0,1]$ one starts in the first level
with the two fractions $m/n=0/1$ and $m'/n'=1/1$.
In each iteration for each pair of adjacent fractions $m/n$ and $m'/n'$ 
we insert the mediant $\zeta=(m+m')/(n+n')$.
This leads to the sequence of sets $\{0/1,1/1\}$, $\{0/1,1/2,1/1\}$, $\{0/1,1/3,1/2,2/3,1/1\},...$.
Alternatively one could also use the Farey tree \cite{HarWri1975, Gut2011} which is a subtree of the Stern--Brocot tree.

2.) For each new rational $\zeta$ of a level one determines
its finite continued fraction expansion
\begin{subequations} \label{eq:ZetaContFrac}
\begin{align} 
  \zeta & = \zeta_0 + \dfrac{1}{\zeta_1 + \dfrac{1}{\zeta_2+\dfrac{1}{\ddots +\dfrac{1}{\zeta_k}}}}, \\
    & =: [\zeta_0; \zeta_1, \zeta_2, ..., \zeta_k].
\end{align}
\end{subequations}
Appending the infinite continued fraction expansion of
the golden mean  $\sigma = (\sqrt{5}-1)/2  =[0; 1,1,1, ...]$
at the end of the continued fraction expansion~\eqref{eq:ZetaContFrac}
gives the noble number
\begin{align}
\label{eq:noble_number}
\xi = [\zeta_0; \zeta_1, \zeta_2, ... , \zeta_k, 1, 1, 1, ...].
\end{align}
The construction is such that there is precisely one such noble number
between each pair of adjacent rationals of a given level of the Stern--Brocot tree.

3.) Each noble number $\xi$ leads to a frequency $\omegaU$ according to Eq.~\eqref{eq:FrequencyXiRelation}.

4.) The iteration is stopped when $\Ntori$ frequencies are found
within the range of frequencies $\omegaU(\crvparam)$ of the regular region.

\end{appendix}

\end{document}